\documentclass[final]{aipproc}

\layoutstyle{8x11double}
\usepackage{xspace} 
\usepackage{graphicx}

\newcommand{\lesssim}{\raisebox{0.3mm}{\em $\, <$} 
\hspace{-2.9mm} \raisebox{-1.5mm}{\em $\sim \,$}}
\newcommand{\gtrsim}{\raisebox{0.3mm}{\em $\, >$} 
\hspace{-2.9mm} \raisebox{-1.5mm}{\em $\sim \,$}}

\begin{document}

\title{Some attempts to explain MINOS anomaly}


\author{Osamu Yasuda}
{address={Department of Physics, Tokyo Metropolitan University,
    Minami-Osawa, Hachioji, Tokyo 192-0397, Japan} }

\begin{abstract}
Some attempts which were made to explain the MINOS anomaly
are critically discussed.  They include 
the non-standard neutral current-neutrino interaction and
the (3+1)-scheme with sterile neutrino.
\end{abstract}

\maketitle

\noindent
{\bf 1. Introduction}
\vskip 0.1in

At the neutrino 2010 conference, the MINOS collaboration
reported that the allowed region for the mass squared difference
obtained from their anti-neutrino data differed from that for
the neutrino data \cite{vahle}.
There have been several attempts to account for this anomaly.
They include the non-standard neutral current-neutrino interaction
with $\mu$, $\tau$ components,
and the (3+1)-scheme with sterile neutrino.
In this talk I will examine whether they are
consistent with other experiments.

\vskip 0.2in
\noindent
{\bf 2. Non-standard interactions in propagation}
\vskip 0.1in

One of the ideas to distinguish neutrinos and
anti-neutrinos is to use the matter effect.
In order to affect $\nu_\mu$ and $\bar{\nu}_\mu$
at the MINOS energy range,
one should introduce the non-standard interaction
in propagation of neutrinos so that
the matter potential has at least non-zero $\mu$
or $\tau$ components.  Here let us consider
a general $3\times3$ potential matrix:
\begin{eqnarray}
A\left(
\begin{array}{ccc}
1+ \epsilon_{ee} & \epsilon_{e\mu} & \epsilon_{e\tau}\\
\epsilon_{\mu e} & \epsilon_{\mu\mu} & \epsilon_{\mu\tau}\\
\epsilon_{\tau e} & \epsilon_{\tau\mu} & \epsilon_{\tau\tau}
\end{array}
\right),
\label{matter-np}
\end{eqnarray}
where $A\equiv\sqrt{2}G_FN_e$
stands for the matter effect.
It was pointed out in Ref.\,\cite{Oki:2010uc}
that with new physics (\ref{matter-np})
the disappearance probability
in the high-energy atmospheric neutrino oscillations
behaves as
\begin{eqnarray}
1-P(\nu_\mu\rightarrow\nu_\mu)\simeq
c_0 + c_1\frac{\Delta m^2_{31}}{AE} + {\cal O}\left(\frac{\Delta m^2_{31}}{AE}\right)^2,
\label{expansion}
\end{eqnarray}
where $c_0$ and $c_1$ are functions of the parameters
$\epsilon_{\alpha\beta}$ of new physics.
On the other hand,
in the standard three-flavor scheme, the
high-energy behavior of the disappearance
oscillation probability is
\begin{eqnarray}
&{\ }&
1-P(\nu_\mu\rightarrow\nu_\mu)
\nonumber\\
&\simeq&
\left(\frac{\Delta m^2_{31}}{2AE}\right)^2\left[
\sin^22\theta_{23}
\left(\frac{c^2_{13}AL}{2}\right)^2
\right.\nonumber\\
&{\ }&\qquad\qquad+\left.
s^2_{23}\sin^22\theta_{13}\sin^2\left(
\frac{AL}{2}\right)
\right],
\label{he-std}
\end{eqnarray}
where the terms of ${\cal O}(1)$
and ${\cal O}(\Delta m^2_{31}/AE)$ are absent in Eq.\,(\ref{he-std})
which is in perfect agreement with the experimental data.
It was shown in Ref.\,\cite{Oki:2010uc} that
$|c_0|\ll1$ and $|c_1|\ll1$ in Eq. (\ref{expansion}) imply
\begin{eqnarray}
|\epsilon_{e\mu}|^2+|\epsilon_{\mu\mu}|^2+|\epsilon_{\mu\tau}|^2\ll1
\label{cond1}\\
||\epsilon_{e\tau}|^2
-\epsilon_{\tau\tau} \left( 1 + \epsilon_{ee} \right)|\ll1,
\label{cond2}
\end{eqnarray}
respectively.\footnote{Eq.\,(\ref{cond2}) was first
found in Ref.\,\cite{Friedland:2004ah}.}

\begin{figure}[htb]
\resizebox{29.9pc}{!}{\includegraphics{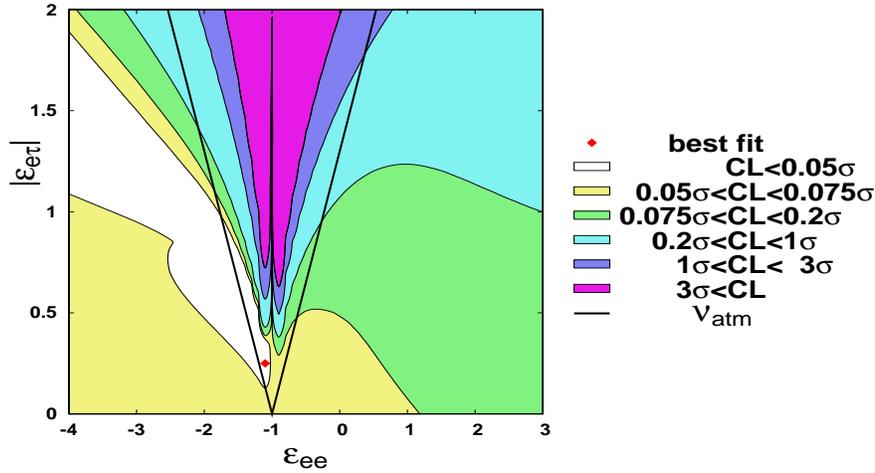}}
\caption{The disfavored region \cite{Yasuda:2010v1}
obtained from the MINOS data \cite{vahle}.
The region above the diagonal straight lines is excluded
because of the atmospheric
neutrino data \cite{Friedland:2006pi}.
The contours are drawn to exaggerate
its significance.
}
\label{fig1}
\end{figure}

\vskip 0.2in
\noindent
{\bf (i) Non-standard interactions in propagation with $\mu$, $\tau$ components}
\vskip 0.1in

The simpler possibility within the ansatz (\ref{matter-np})
is to assume that all the electron components $\epsilon_{e\alpha}$ vanish:
\begin{eqnarray}
A\left(
\begin{array}{ccc}
1 & 0 & 0\\
0 & \epsilon_{\mu\mu} & \epsilon_{\mu\tau}\\
0 & \epsilon_{\tau\mu} & \epsilon_{\tau\tau}
\end{array}
\right).
\label{matter-mt}
\end{eqnarray}
In this case, since the contribution from the solar neutrino
oscillation is negligible for the range of the energy and
the baseline length of MINOS, $\nu_e$ decouples from
$\nu_\mu$ and $\nu_\tau$.  Refs.\,\cite{Mann:2010jz}
and \cite{Kopp:2010qt} performed an analysis with the
ansatz (\ref{matter-mt}), where
$\epsilon_{\mu\mu}=\epsilon_{\tau\tau}=0$ was assumed in the former work.
The best fit values for
$\epsilon_{\mu\tau}$ obtained
in Refs.\,\cite{Mann:2010jz,Kopp:2010qt}
do not satisfy the constraint from the atmospheric neutrino data
$|\epsilon_{\mu\tau}|\lesssim 7\times 10^{-2}$ at 90\%CL
\cite{Fornengo:2001pm,GonzalezGarcia:2004wg,Mitsuka:2008zz},
so their solutions are inconsistent with
the atmospheric neutrinos.\footnote{
The two flavor ansatz (\ref{matter-mt}) can be regarded as a subset of
the three flavor scenario in the limiting case
$\epsilon_{ee}=\epsilon_{e\mu}=\epsilon_{e\tau}=\theta_{13}=\Delta m_{21}^2=0$, so
the constraint (\ref{cond2}) in the two flavor case leads to
$|\epsilon_{\tau\tau}|\simeq0$.
On the other hand, the bound on $|\epsilon_{\mu\tau}|$
in the three flavor case is independent of other components
$\epsilon_{\alpha\beta}$, so
the bound $|\epsilon_{\mu\tau}|\lesssim {\cal O}(10^{-2})$
in Refs.\,\cite{Fornengo:2001pm,GonzalezGarcia:2004wg,Mitsuka:2008zz}
is expected to be valid both in the two and three flavor cases.}

\vskip 0.2in
\noindent
{\bf (ii) A model with gauging $L_\alpha-L_\beta$}
\vskip 0.1in

Ref.\,\cite{Heeck:2010pg} discussed the model
with gauging the lepton numbers $L_\alpha-L_\beta$.
Such models predict the matter potentials
$\mbox{\rm diag}(V,-V,0)$,
$\mbox{\rm diag}(V,0,-V)$,
and $\mbox{\rm diag}(0,V,-V)$
for $L_e-L_\mu, ~L_e-L_\tau, ~L_\mu-L_\tau$,
respectively,
where the major contribution to the potential $V$ 
comes from the Sun instead of the matter in the Earth.
In order for this scenario to account for the MINOS
anomaly, the matter effect $V$ should be
comparable to $|\Delta m^2_{31}|/E_\nu^{\mbox{\rm\tiny MINOS}}$ in
magnitude.  On the other hand, since the matter effect $V$
mainly comes from the Sun, if $\alpha$ or
$\beta$ in $L_\alpha-L_\beta$ is of electron type,
then the magnitude of $V$
for the solar neutrino oscillation
is expected to be enhanced by the factor
(distance between Sun and Earth)/(radius of Sun),
and it would destroy the success of the
oscillation interpretation of the solar neutrino deficit,
because its matter effect would be much
larger than the standard one.
To avoid its influence on the solar neutrino oscillation,
one is forced to work with $L_\mu-L_\tau$.
In this case, however, it would contradict
with the atmospheric neutrino
constraint $|\epsilon_{\mu\mu}-\epsilon_{\tau\tau}|\ll1$
\cite{Fornengo:2001pm,GonzalezGarcia:2004wg,Mitsuka:2008zz}.
So all the channels have conflict with one
experiment or the other.

\vskip 0.2in
\noindent
{\bf (iii) Non-standard interactions in propagation with $e$, $\tau$ components}
\vskip 0.1in

Taking into account the constraint from the
atmospheric neutrino data, the only possibility
which could potentially produce large difference
between neutrinos and anti-neutrinos is the form
of the potential:
\begin{eqnarray}
{\cal A}= A\left(
\begin{array}{ccc}
1+ \epsilon_{ee}~~ & 0 & \epsilon_{e\tau}\\
0 & 0 & 0\\
\epsilon_{e\tau}^\ast & 0 & ~~|\epsilon_{e\tau}|^2/(1 + \epsilon_{ee})
\end{array}
\right),
\label{matter-et}
\end{eqnarray}
where $|\epsilon_{ee}|\lesssim4$,
$\epsilon_{e\tau}\lesssim3$ are allowed
at 90\%CL from all the experimental data
(see Ref.\,\cite{Oki:2010uc,Yasuda:2010hw} and
references therein).
Although this potential term does not have mixing
between $\nu_\mu$ and $\nu_e$ or $\nu_\tau$,
it can affect $\nu_\mu$ through the
(maximal) mixing between $\nu_\mu$ and
$\nu_\tau$ in vacuum.
The region in which the MINOS anomaly can be
accounted for by the ansatz (\ref{matter-et})
is given in Fig.\ref{fig1} \cite{Yasuda:2010v1}.
This result has two undesirable features.
Firstly, the best fit point lies in
the region which is excluded by
the atmospheric neutrino data \cite{Friedland:2006pi}.
Secondly, while the disfavored region at 3$\sigma$CL
almost coincides with the one by the atmospheric
neutrino data \cite{Friedland:2006pi},
the significance of the standard case
($\epsilon_{ee}=\epsilon_{e\tau}=0$)
compared with the best fit point
is only 0.07$\sigma$CL.
Therefore, we conclude that it is not
worth introducing this scenario to
explain the MINOS anomaly.

\vskip 0.2in
\noindent
{\bf 3. A (3+1)-scheme with one sterile neutrino ($\nu_s$)}
\vskip 0.1in

The other scenario I would like to discuss is the (3+1)-scheme
with one sterile neutrino.
Here let us take the parametrization \cite{Donini:2007yf}
\begin{eqnarray}
    U &=&
    R_{34}(\theta_{34} ,\, 0) \; R_{24}(\theta_{24} ,\, 0) \;
    R_{23}(\theta_{23} ,\, \delta_3) \nonumber\\
&{\,}&\times
    R_{14}(\theta_{14} ,\, 0) \; R_{13}(\theta_{13} ,\, \delta_2) \; 
    R_{12}(\theta_{12} ,\, \delta_1),
\label{eq:3+1param2}
\end{eqnarray}
where
$\left[R_{ij}(\theta ,\, \delta)\right]_{pq}\equiv
\delta_{pq}+(\cos\theta-1)
(\delta_{pi}\delta_{qi}+\delta_{pj}\delta_{qj})
+\sin\theta(e^{-i\delta}\delta_{pi}\delta_{qj}
-e^{i\delta}\delta_{pj}\delta_{qi}
)$
is a $4\times 4$ rotational matrix 
which mixes $i$ and $j$ components with a mixing
angle $\theta$ and a CP phase $\delta$.
It is known
that $\sin^22\theta_{13}\ll 1$, $\sin^22\theta_{14}\ll 1$
should follow from the constraints of
the reactor experiments \cite{Declais:1994su,Apollonio:1999ae},
and, 
if $0.7\mbox{\rm eV}^2\lesssim\Delta m^2_{41}\lesssim 10\mbox{\rm eV}^2$,
$\sin^22\theta_{24}\lesssim 0.2$ should hold
to satisfy the constraint of the CDHSW experiment \cite{Dydak:1983zq}.
Furthermore, one can show that the coefficient $c_0$ in
the high energy behavior (\ref{expansion})
is proportional to $\sin^22\theta_{24}$, so
$\theta_{24}$ should be small also from
the atmospheric neutrino constraint.\footnote{
According to the analysis in Ref.\,\cite{Donini:2007yf},
the allowed region for $\theta_{24}$ at 90\%CL
is $0\le\theta_{24}\lesssim ~\pi/15$.}
Here for simplicity I assume $\theta_{13}=\theta_{14}=\theta_{24}=0$
to be consistent with the constraints from the reactor, CDHSW
and atmospheric neutrino data.
In this case, $\nu_e$ decouples from $\nu_\mu$, $\nu_\tau$ and
$\nu_s$, and the situation becomes similar to that
of the solar neutrino oscillations in the standard case.
The disappearance probability in this case is given by
\begin{eqnarray}
&{\ }&
\left\{
\begin{array}{c}
1-P(\nu_\mu\rightarrow\nu_\mu)\\
1-P(\bar{\nu}_\mu\rightarrow\bar{\nu}_\mu)
\end{array}
\right\}
\nonumber\\
&\sim&
\left(\frac{\Delta E_{32}}{\Delta\tilde{E}_{32}^{(\pm)}}
\right)^2\sin^22\theta_{23}
\sin^2\left(\frac{\Delta\tilde{E}_{32}^{(\pm)}L}{2}\right)\qquad
\label{disappearance-s}\\
&{\ }&
\Delta\tilde{E}_{32}^{(\pm)}\equiv
\left[(\Delta E_{32}\cos2\theta_{23}\pm \sin\theta^2_{34}A/2)^2
\right.
\nonumber\\
&{\ }&\qquad\quad\left.+(\Delta E_{32}\sin2\theta_{23})^2\right]^{1/2},
\nonumber
\end{eqnarray}
where $\Delta E_{32}\equiv\Delta m^2_{32}/2E$ and
small quantities such as $\Delta m^2_{32}/\Delta m^2_{42}$
have been ignored.
$\theta_{34}$ stands for the mixing angle
which represents the ratio of
$\nu_\mu\leftrightarrow\nu_\tau$ and
$\nu_\mu\leftrightarrow\nu_s$
oscillations, and
deviation of Eq.\,(\ref{disappearance-s}) from the oscillation probability
in vacuum becomes larger
as $\theta_{34}$ increases.  The
matter effect becomes important for
the energy range $E\gtrsim 10$GeV,
so the zenith angle dependence of
the high-energy atmospheric data
gives a constraint on $\theta_{34}$.
The analysis in Ref.\,\cite{Donini:2007yf}
tells us that the allowed region at 90\%CL
by the atmospheric neutrino data is
$0\le \theta_{34}\lesssim \,\pi/6$.
Eq.\,(\ref{disappearance-s}) is potentially interesting
because non-zero $\theta_{34}$ distinguishes
the effective mixing angles and the effective
mass squared differences of neutrinos and anti-neutrinos.
However, because the atmospheric mixing angle
$\theta_{23}$ is nearly maximal ($|\cos2\theta_{23}|\ll1$),
it is difficult in practice to distinguish
neutrinos and anti-neutrinos from Eq.\,(\ref{disappearance-s}).
In fact, according to the
numerical analysis \cite{Yasuda:2010v1},
the best fit point with the present (3+1)-scheme
is the same as that for the standard case.\footnote{
Ref.\,\cite{Engelhardt:2010dx} performed a similar analysis
using the old MINOS data,
but they obtained a result different from ours.
}
Also in this case, therefore, it is difficult
to explain the MINOS
anomaly.\footnote{The situation of the
interpretation as sterile neutrino oscillations to account for
the LSND anomaly is still confusing because of the
MiniBooNE anti-neutrino data \cite{AguilarArevalo:2010wv}.
The (3+1)-scheme which I discussed here
predict null results for the $\nu_\mu\to\nu_e$
and $\bar{\nu}_\mu\to\bar{\nu}_e$ channels
at the $L/E$ range of the LSND and MiniBooNE experiments,
because $P(\nu_\mu\to\nu_e)=P(\bar{\nu}_\mu\to\bar{\nu}_e)
=\sin^2\theta_{24}\sin^22\theta_{14}\sin^2(\Delta m^2_{41}L/4E)=0$
in the present assumption.  If the LSND anomaly is real, we can
take small mixing angles $\theta_{14}$ and $\theta_{24}$
into account within the framework of the present (3+1)-scheme.
Even in that case, however, the effect of these
mixing angles on the disappearance channels $\nu_\mu\to\nu_\mu$
and $\bar{\nu}_\mu\to\bar{\nu}_\mu$ is small
and the present conclusion
does not change.}

\vskip 0.2in
\noindent
{\bf 4. Conclusion}
\vskip 0.1in

Unfortunately, none of the scenarios, which have been proposed
so far to explain the MINOS anomaly, seem to work.  They either
give little contribution to distinguish neutrinos and
anti-neutrinos, or excluded by the constraints of other experiments.
Since the MINOS anomaly is only a $2\sigma$ effect,
probably we should wait until we have more statistics.

\vskip 0.2in
\noindent
{\bf Acknowledgement}
\vskip 0.1in

I would like to thank the organizers for
invitation and hospitality during the workshop.
I would like to thank Thomas Schwetz for his help
in reproducing the results by MINOS.
This research was partly supported by a Grant-in-Aid for Scientific
Research of the Ministry of Education, Science and Culture, under
Grant No. 21540274.

\end{document}